\documentclass[twocolumn,prl,aps,floatfix,superscriptaddress,amssymb]{revtex4}
\usepackage{epsfig}
\usepackage{bm}

\def\bc{\begin{center}}
\def\ec{\end{center}}
\def\be{\begin{equation}}
\def\ee{\end{equation}}
\def\bea{\begin{eqnarray}}
\def\eea{\end{eqnarray}}

\begin{document}

\title{Interaction-Induced Strong Localization in Quantum Dots}

\author{A.~D.~G\"u\c{c}l\"u}
\affiliation{Department of Physics, Duke University, Box 90305,
Durham, North Carolina 27708-0305}

\author{Amit~Ghosal}
\affiliation{Physics Department, University of California Los Angeles, Los
Angeles, California 90095-1547}

\author{C.~J.~Umrigar}
\affiliation{Laboratory of Atomic and Solid State Physics, Cornell University,
Ithaca, New York 14853}

\author{Harold~U.~Baranger}
\affiliation{Department of Physics, Duke University, Box 90305,
Durham, North Carolina 27708-0305}

\date{October 3, 2007; published as Phys. Rev. B 77, 041301(R) (2008)}

\begin{abstract}

We argue that Coulomb blockade phenomena are a useful probe of the
cross-over to strong correlation in quantum dots. Through calculations
at low density using variational and diffusion quantum Monte Carlo (up
to $r_s\!\sim\!55$), we find that the addition energy shows a clear
progression from features associated with shell structure to those
caused by commensurability of a Wigner crystal. This cross-over (which
occurs near $r_s\!\sim\!20$ for spin-polarized electrons) is, then, a
signature of interaction-driven localization. As the addition energy
is directly measurable in Coulomb blockade conductance experiments,
this provides a direct probe of localization in the low density
electron gas.
\end{abstract}

\maketitle



Localization of electrons induced by electron-electron interactions is a key issue in strongly-interacting systems \cite{KivelsonFradkinRev05} which dates from the dawn of solid state physics when Wigner introduced the notion of an electron crystal \cite{Wigner34}. The recent focus on localization in inhomogeneous systems is stimulated in part by experiments which apparently see a metal-insulator transition in two dimensions \cite{KravSarachik04}. This occurs in low-density electron gas samples in which disorder is presumably important. Recall that in a degenerate electron gas, low density implies strong interactions. Here we show that an approach from nanoscale physics, in which the potential confining the electrons plays a key role analogous to that of disorder, provides new information on this critical problem.


The Coulomb blockade effect has been a valuable tool for probing a variety of interaction effects in quantum dots \cite{HeissQdotBook,ReimannMannRMP02,KouTaruchaRPP01,DavidGGKondoRev07}. By ``quantum dot'' we mean a confined region of electron gas containing between $N=1$ and $\sim\! 1000$ electrons; experimentally, they have proven remarkably tunable through the use of gate voltages on nearby electrodes. We treat $N \le 20$ here. The large electrostatic charging energy usually forces the dot to have a fixed number of electrons, preventing the flow of current. The blockade is lifted by using a gate to tune the energy for $N$ electrons to be the same as that for $N+1$, inducing a finite conductance. The conductance through the dot as a function of gate voltage is therefore a series of sharp peaks. The height and position of these peaks encode information about the dot's ground state; for instance, the spacing between the peaks is proportional to the second difference of the ground-state energy with respect to electron number $N$, a quantity known as the addition energy. Quantum many-body physics probed in quantum dots includes, for instance, the atomic-like effect of exchange and correlation in altering the filling of single-particle shells \cite{KouTaruchaRPP01}, several kinds of Kondo effects \cite{DavidGGKondoRev07}, aspects of the fractional quantum Hall effect \cite{ReimannMannRMP02}, and the entanglement of spin and orbital degrees of freedom for quantum information purposes \cite{HeissQdotBook}.



We show that the Coulomb blockade can be a valuable probe of
localization in quantum dots. In particular, the characteristic
pattern of the addition energy changes as the dot crosses over from
the extended states of the high density Fermi liquid regime to the
localized states characterizing the low density Wigner
crystal. Interaction strength is often characterized by
$r_s=1/a_B^*\sqrt{\pi n}$ (in two dimensions) where $a_B^*$ is the
effective Bohr radius and $n$ is the electron density. 
In our quantum dots, the cross-over occurs for $r_s$ substantially smaller than the value \cite{TC89,Attaccalite02,Waintal06} in bulk. This decrease in interaction strength needed for localization is connected to the density inhomogeneity necessarily present in this confined system; density inhomogeneity produced in other ways, such as by disorder, may similarly enhance localization. 
Our main point is that such a cross-over to localization can be directly measured in Coulomb blockade experiments. 


The cross-over from Fermi liquid to ``Wigner molecule'' in quantum dots was studied previously using various many-body methods. Exact diagonalization \cite{ReimannMannRMP02,RCB+06}, while the most robust and direct approach, is limited to small electron number and small $r_s$ due to convergence problems. Path-integral quantum Monte Carlo \cite{Egger99,FBL01,WE05} is well suited for finite temperature properties, but preserves only $S_z$ symmetry and has large statistical fluctuations at low temperatures. Variational and diffusion Monte Carlo \cite{FMN+01} preserve $S_z$, $S^2$, and $L_z$, symmetry;  though limited by the ``fixed node'' error which depends on the quality of the trial wave function \cite{FMN+01}, they were successfully used to study quantum dots for $r_s$ up to $\sim\!4$ \cite{PederivaUmrigar00,HSN02,GJU+05,GUJ+05} and then up to $r_s\sim\!15$ in our recent work \cite{GGU+06,GGU+06b}. Here, we apply recently developed energy minimization methods \cite{UmrigarFilippi05,Umrigar07} to floating gaussian-based trial wave functions, enabling us to decrease the fixed-node error and investigate the strongly correlated regime up to as high as $r_s\sim\!55$.


Our model quantum dot consists of $N$ interacting electrons in a two-dimensional circular quadratic potential. The Hamiltonian, expressed in effective atomic units (electronic charge $e$, dielectric constant $\epsilon$, effective mass $m^*$, and $\hbar$ are set to $1$), is given by
\begin{eqnarray}
H=-\frac{1}{2}\sum_i^N\bigtriangledown_i^2
 +\frac{1}{2}\sum_i^N\omega^2 r_i^2
 +\sum_{i < j}^N {1 \over r_{ij}}
\label{Hamilton}
\end{eqnarray}
where $\omega$ is the spring constant of the quadratic potential which provides control of the strength of the Coulomb interaction with respect to the kinetic energy. In analogy with 2D bulk systems, we characterize the interaction strength using $r_s=(\pi \bar{n})^{-1/2}$, where $\bar{n}\equiv \int n^2({\bf r})d{\bf r}/N$ is the mean density of electrons.

Variational (VMC) and diffusion (DMC) Monte Carlo techniques \cite{FMN+01}
were used to calculate the properties of our model quantum dots.
One starts with a set of single-particle orbitals---simple gaussian functions or from a self-consistent calculation (Hartree or Kohn-Sham). We then perform a VMC calculation using a trial wave function, $\Psi_T$, which is a linear combination of products of up- and down- spin Slater determinants of these orbitals multiplied by a Jastrow factor.  (The detailed form of our Jastrow factor is in Ref.~\onlinecite{GJU+05}.) In a second stage, we use fixed-node DMC \cite{FMN+01,UmrigarNightingaleRunge93} to project the optimized many-body wave function onto a better approximation of the true ground state, an approximation that has the same nodes as $\Psi_T$. The fixed-node DMC energy is an upper bound to the true energy and depends only on the nodes of the trial wave function obtained from VMC.

In order to capture the Wigner molecule aspect, we build our trial wave functions from floating gaussian orbitals. The use of localized orbitals in confined structures causes three problems: (i) The positions of the gaussians are initially unknown and must be optimized together with the widths. (ii) Slater determinants built from localized gaussians do not conserve angular momentum unless integrated over all angles \cite{LYL06}. Strikingly, we find that taking a linear superposition of as few as $3$ rotated determinants (for each ring) recovers more than $99\%$ of the missing DMC energy. 
(iii) Low spin states require the superposition of many Slater determinants in order to have an exact spin eigenstate, making spin polarized states easier to study. All the parameters in our VMC calculations---Jastrow parameters, widths and positions of the gaussians, and determinantal coefficients---are optimized using efficient energy minimization techniques \cite{UmrigarFilippi05,Umrigar07}. The use of these techniques is a key feature which ultimately allows us to cover a much wider range of density with this method than was previously possible.

We have also used liquid-like orbitals obtained by solving Kohn-Sham or Hartree equations. A comparison of the results obtained from different types of orbitals is presented below. Quantities that do not commute with the Hamiltonian were calculated using an extrapolated estimator $F_{\rm  QMC}=2F_{\rm DMC}-F_{\rm VMC}$ \cite{FMN+01}.

\begin{figure}
\epsfig{file=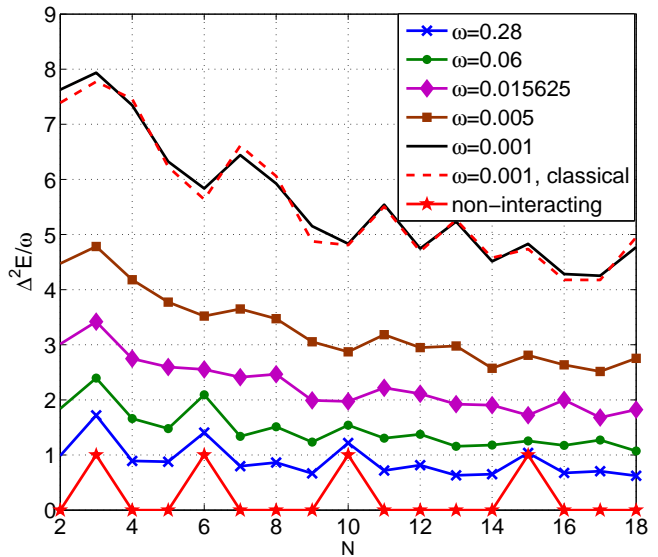,width=3.375in}
\caption{\label{fig:addn_fullypol} Addition energy as a function of number of electrons in the quantum dot for different
$\omega$ values (corresponding $r_s$ values are approximately
0, 2, 5, 15, 25, 55). The classical \cite{BP94} and non-interacting addition
energies are also shown. Note the progression from peaks determined by shell structure to those consistent with classical electrostatics.}
\end{figure}


In Fig.~\ref{fig:addn_fullypol} we plot the addition energy
$\Delta^2E(N)=E_G(N+1)+E_G(N-1)-2E_G(N)$ of the fully spin polarized
ground states for different $\omega$ values, including the results for
the classical \cite{BP94} and non-interacting systems. As the addition
energy is directly measurable in Coulomb blockade conductance
experiments \cite{ReimannMannRMP02,KouTaruchaRPP01}, the structure of
the curves in Fig.~\ref{fig:addn_fullypol} can be observed by applying
a strong in-plane magnetic field. In the non-interacting case
($\omega\rightarrow\infty$), the peaks at $N=3$, $6$, $10$, and $15$
are due to the shell structure of the two-dimensional harmonic
confinement potential. As $\omega$ is decreased, the electronic
density decreases and electron-electron interactions become more
important. Note that this causes the strength of the fluctuations in
$\Delta^2E$ to decrease. Around $\omega\sim0.015$ ($r_s\sim15$), the
shell structure peaks are washed out by the Coulomb interaction
energy. For larger $r_s$ values, a new structure sets in, showing
peaks at $N=3$, $7$, $11$, $13$, $15$, and $18$. These new magic
numbers are caused by commensurability of a Wigner crystal. In fact,
at $r_s\sim 55$, there is an impressive quantitative agreement between
the quantum mechanical and classical addition energies. \textit{Our
results suggest that the cross-over from a liquid-like state towards
the Wigner crystal regime in quantum dots, which occurs around
$r_s\sim 20$ for fully polarized electrons, can be directly observed
experimentally.}

\begin{figure}
\epsfig{file=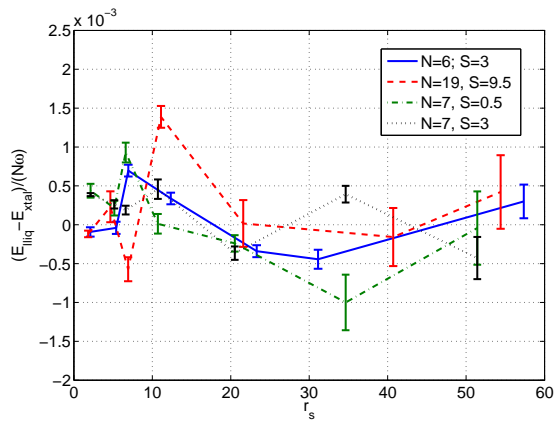,width=0.45\textwidth}
\caption{\label{fig:liq_vs_xtal} Difference between the DMC energies obtained using liquid-like (Hartree) orbitals and solid-like (localized gaussian) orbitals as a function of $r_s$. Both types of orbitals appear to work well throughout the whole range of $r_s$.}
\end{figure}

In the 2D electron gas, QMC methods were previously used to study the liquid-solid transition by comparing the fixed-node DMC energies as a function of $r_s$ obtained using liquid-like orbitals and localized gaussian orbitals \cite{TC89,Attaccalite02,Waintal06}. The energies using the two types of orbitals cross near $r_s\sim 35$, which was therefore identified as the localization point. In order to investigate if such a cross-over occurs in quantum dots, we compared the DMC energies obtained using delocalized Hartree orbitals and localized gaussian orbitals for several $(N,S)$ states. Fig.~\ref{fig:liq_vs_xtal} shows the energy differences for 4 different cases. To our surprise, we did not find any systematic $r_s$ dependence: solid and liquid type orbitals give equally good DMC energies at $r_s$ values ranging from $\sim\!2$ to $55$.

We now consider an alternative way of quantifying the degree of
localization as a function of $r_s$:
We define the angular power spectrum of the quantum dot as
\begin{eqnarray}
  f(r,k_\theta)&\equiv&\int d{\bf r}_1..d{\bf r}_N \vert \Psi({\bf r}_1,..{\bf r}_N)\vert^2
 \nonumber \\
  & &\times\, \Big\vert \sum_i^N \mathcal{F}_\theta \lbrace \delta^2({\bf r}_i-{\bf r}) \rbrace \Big\vert^2 -\frac{n(r)}{r^2}
 \nonumber \\
  &=& \frac{2}{r^2}\int d{\bf r}_1..d{\bf r}_N \vert \Psi({\bf r}_1,..{\bf r}_N)\vert^2
\label{powspec} \\
  & &\times  \sum_{i>j}^N \cos[k_\theta (\theta_i-\theta_j)] \delta(r_i-r)\delta(r_j-r)
 \nonumber
\end{eqnarray}
where ${\bf r}=(r,\theta)$ is the position vector and $\mathcal{F}_\theta$ is the Fourier transform in the angular direction. For a classical ring of $N_r$ equally spaced electrons, the above expression is proportional to $\sum_{m=1}^{N_r} (N_r -m) \cos(2 \pi m k_\theta/N_r)$, which has maxima at $k_\theta=0$, $N_r$, etc.

\begin{figure}
\epsfig{file=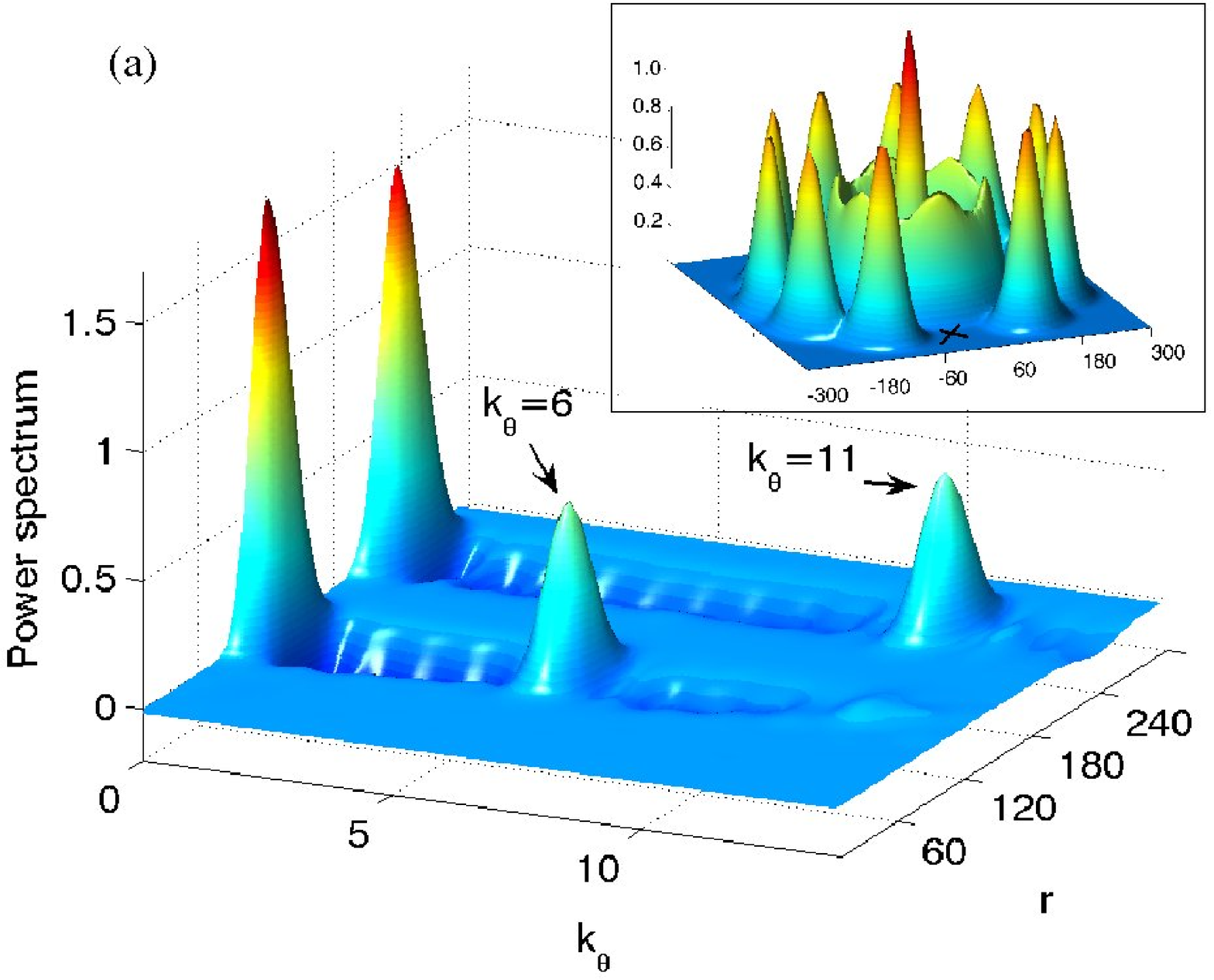,width=3.375in}
\epsfig{file=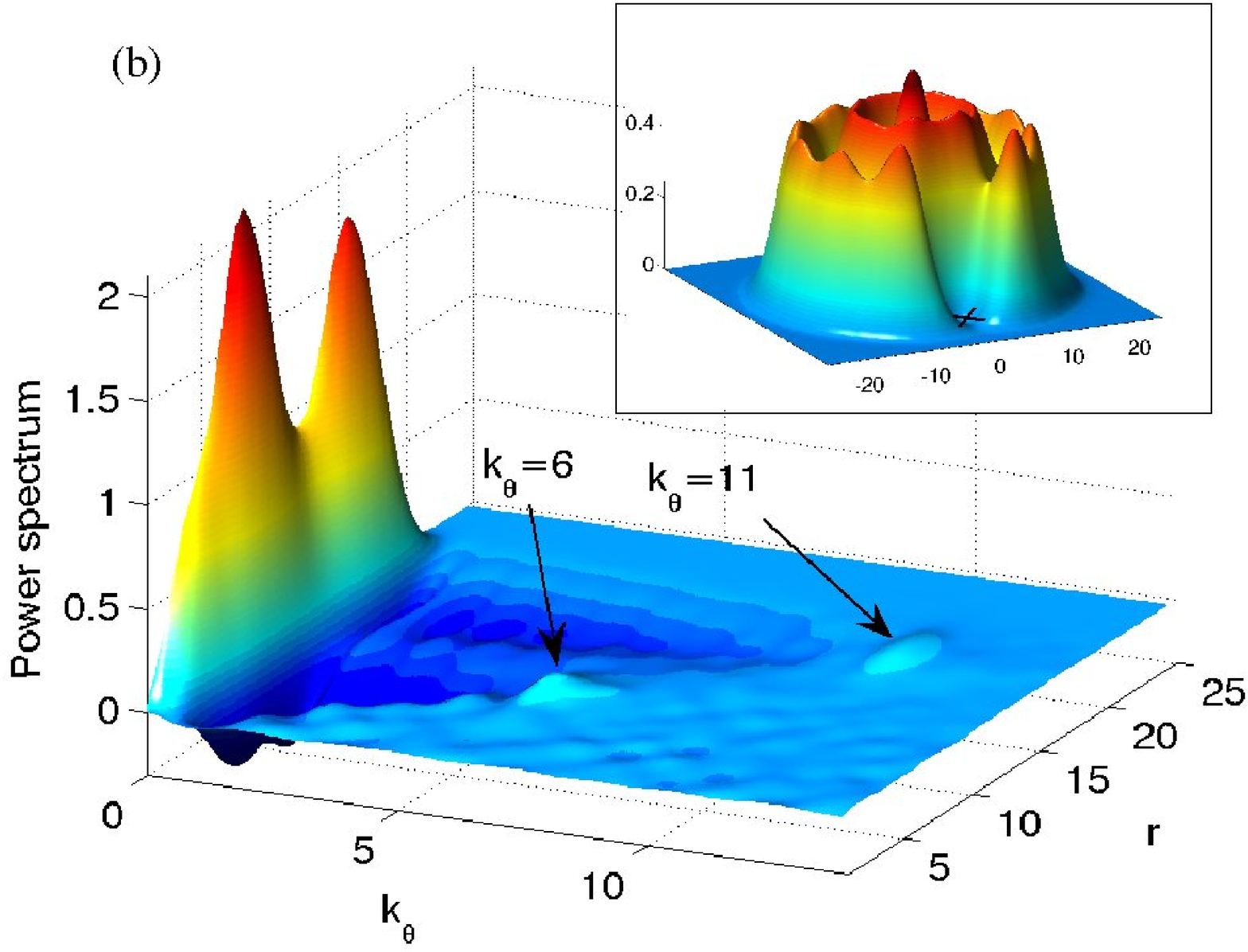,width=3.375in}
\caption{\label{fig:ps+pd}
Angular power spectrum for $N=18$ as a function of $k_\theta$ and quantum dot radius $r$ for (a) $r_s=52$ and (b) $r_s=4.8$. Insets: Corresponding pair-densities (fixed electron marked by ``x''). From the peaks in the power spectrum, one deduces the number of electrons in each ring and the strength of the angular modulation.
}
\end{figure}

Fig.~\ref{fig:ps+pd}(a) shows the power spectrum of a fully polarized system of $N=18$ electrons at $r_s=52$. From this plot, three important quantitative points can be extracted: (i)~At $k_\theta=0$, Eq.~(\ref{powspec}) reduces to the probability of finding two electrons at a given radius; thus, the peaks at $k_\theta=0$ give the radial location of the electrons. There is an inner ring at $r\approx110$ well separated from the outer ring at $r\approx230$ (plus a single electron at the center). (ii) There are two more peaks at $k_\theta=6$ and $k_\theta=11$, indicating the localization of $6$ and $11$ electrons in the inner and outer rings, respectively. (iii) The height of the secondary peaks compared to the $k_\theta=0$ peaks is a measure of the angular localization of the electrons in the rings. For comparison, the pair-density is shown in the inset---the density of electrons at ${\bf r}$ given that one electron is fixed at the position marked with the cross. We note that the (1,6,11) configuration (that is, 1 electron at the center, 6 in the inner ring, and 11 in the outer ring) indicated by both the power spectrum and pair-density plots also corresponds to the classical configuration \cite{BP94}. Clearly, at $r_s\sim52$, the electrons are well localized at their classical position, confirming the agreement between quantum mechanical and classical addition energies in Fig.~\ref{fig:addn_fullypol}.

On the other hand, Fig.~\ref{fig:ps+pd}(b) shows the results for $r_s\approx5$. The outer and inner rings are not separated as well, and the peaks at $k_\theta=6$ and $k_\theta=11$ are much weaker, although still visible despite the relatively low $r_s$.
The pair-density shows weak intrashell angular modulation consistent with the $k_\theta=11$ peak and almost no intershell angular correlation.

%

\begin{figure}
\epsfig{file=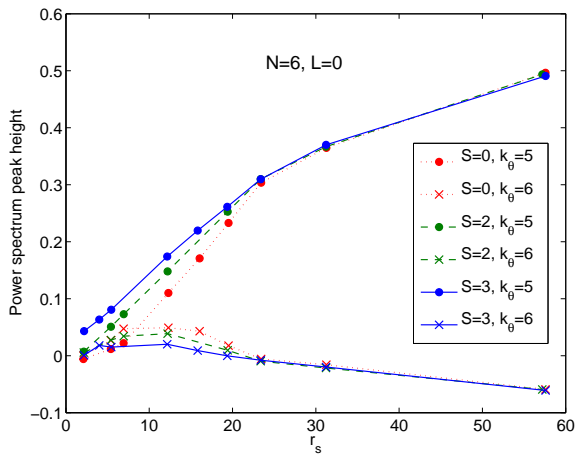,width=3.375in}
\caption{\label{fig:peak_S0-3} The height of the peaks in the angular power spectrum at $r=r_{\rm max}$ for 6 electrons in three cases: fully polarized ($S=3$), unpolarized, and $S=2$ partially polarized. For $S=0$, the number of electrons in the outer ring changes from 6 to 5 as the strength of the interactions increases.}
\end{figure}

For small particle number, $N \le 8$, we studied several spin states of the quantum dot, not just the fully polarized case. $N=6$ is particularly interesting: two spatial configurations of the electrons have similar energies---there are either 6 electrons in one ring or 5 electrons in a ring around 1 electron. Fig.~\ref{fig:peak_S0-3} shows the development of angular modulation in this case. For full polarization, the amplitude of the $(1,5)$ configuration smoothly increases. We find this type of smooth increase for all values of $N$ when fully polarized. In contrast, for $S=0$, $(0,6)$ is dominant at small $r_s$ while $(1,5)$ takes over for strong interaction. For $r_s \agt 20$, the amplitude of the modulation is identical in the two cases, as well as for the intermediate $S=2$ case. This intriguing insensitivity of the spatial behavior to the spin state is another indication of localization.



In conclusion, we have proposed a new way of probing Wigner
localization in quantum dots and presented three kinds of evidence for a
cross-over from extended to localized electronic states at around $r_s
\!\sim\! 20$. 
First, the
angular power spectrum develops sharp peaks indicating that the
positions of the electrons become highly correlated
(Fig.~\ref{fig:ps+pd}). Second, for $N=6$ in which two spatial
configurations of the electrons compete, the angular modulation
becomes completely independent of the spin state
(Fig.~\ref{fig:peak_S0-3}). Finally and most importantly, structure in
the addition energy for spin-polarized electrons as a function of
electron number shows a clear progression from peaks consistent with
shell structure to those consistent with the electrostatics of point
particles (Fig.~\ref{fig:addn_fullypol}). This latter effect should be
observable in Coulomb blockade conductance measurements, allowing
for the first time a direct probe of electronic localization in a 
two-dimensional system.

We thank D. Ullmo and V. Elser for valuable conversations. This work was supported in part by the NSF (DMR-0506953) and by the DOE-CMSN (DE-FG02-07ER46365) grant. AG was supported in part by funds from the David Saxon Chair at UCLA.


\bibliography{bibi1.bib,qdotqmc.bib}

\end{document}